\documentclass[twocolumn,epjc3upd]{svjour3}          %twocolumn
\RequirePackage[T1]{fontenc}
\smartqed  % flush right qed marks, e.g. at end of proof
\RequirePackage{graphicx}
\RequirePackage{mathptmx}      % use Times fonts if available on your TeX system
\RequirePackage{flushend}
\RequirePackage[numbers,sort&compress]{natbib}
\RequirePackage[colorlinks,citecolor=blue,urlcolor=blue,linkcolor=blue]{hyperref}
\usepackage{array}
\usepackage{arydshln}
\usepackage{textcomp}
\journalname{Eur. Phys. J. C}
\usepackage[switch]{lineno}

\usepackage{pgf}                % Package defining color names
\usepackage[normalem]{ulem}   % strikethrough
\usepackage{upgreek}
\usepackage{xspace}             % Spaces that don't get eaten
\usepackage{xfrac}
\usepackage{units}
\usepackage{tabularx}
\usepackage{multirow}
\usepackage{amsmath}
\usepackage{doi}
\usepackage{graphics}

\newcommand{\otwo}{\ensuremath{\mathrm{O}_2}\xspace}
\newcommand{\Tlpm}{\ensuremath{\tau_{\mathrm{C}}}\xspace}
\newcommand{\Tdist}{\ensuremath{\tau_\mathrm{S}}\xspace}
\newcommand{\Tcl}{\ensuremath{\tau_{\mathrm{C}}}\xspace}
\newcommand{\xpb}{\ensuremath{x_\mathrm{P}}\xspace}
\newcommand{\npb}{\ensuremath{n_\mathrm{P}}\xspace}
\newcommand{\xpm}{\ensuremath{x}\xspace}
\newcommand{\npm}{\ensuremath{n}\xspace}
\newcommand{\xl}{\ensuremath{x}\xspace}
\newcommand{\xg}{\ensuremath{x_\mathrm{G}}\xspace}
\newcommand{\Vl}{\ensuremath{V}\xspace}
\newcommand{\Lpb}{\ensuremath{\Lambda_{\mathrm{P}}}\xspace}
\newcommand{\Linj}{\ensuremath{\Lambda_{\mathrm{I}}}\xspace}
\newcommand{\rhog}{\ensuremath{\rho_\mathrm{G}}\xspace}
\newcommand{\rhol}{\ensuremath{\rho_\mathrm{L}}\xspace}
\newcommand{\Asurf}{\ensuremath{A\xspace}}
\newcommand{\ndotsurf}{\ensuremath{\dot{n}_\mathrm{\otwo}}\xspace}
\newcommand{\xdotsurf}{\ensuremath{\dot{x}_\mathrm{S}}\xspace}
\newcommand{\hm}{\ensuremath{h}\xspace}

\newcommand{\eff}{\ensuremath{\epsilon}\xspace}

\newcommand{\deltaP}{\ensuremath{\Delta P_{\mathrm{pump}}}\xspace}
\newcommand{\Csurf}{\ensuremath{C_{\mathrm{S}}}\xspace}
\newcommand{\Cbl}{\ensuremath{C_{\mathrm{BL}}}\xspace}
\newcommand{\rn}{${}^{222}$Rn\xspace}
\newcommand{\ra}{${}^{226}$Ra\xspace}
\newcommand{\mmxe}{\ensuremath{M_\mathrm{Xe}}\xspace}
\newcommand{\ml}{\ensuremath{m}\xspace}
\newcommand{\mdot}{\ensuremath{\dot{m}}\xspace}

\usepackage[separate-uncertainty=true, per-mode=repeated-symbol, allow-number-unit-breaks=true]{siunitx}
\DeclareSIUnit\slpm{\text{SLPM}}
\DeclareSIUnit\tonne{\text{tonne}}
\DeclareSIUnit\bara{\text{bar(a)}}
\DeclareSIUnit\molarity{\text{M}}

\newcommand{\henry}{\SI{62.5}{}\xspace}
\newcommand{\PMcharlen}{\SI{36.5}{\milli \meter}\xspace}
\newcommand{\PMsherwood}{\SI{0.46}{}\xspace}
\newcommand{\PMh}{\SI{1.61e-4}{\m \per \second}\xspace}
\newcommand{\PBhprediction}{\SI{1.49e-4}{\m \per \second}\xspace}
\newcommand{\PBtdistcalcval}{\SI{0.55}{\hour}\xspace}
\newcommand{\PMotwofluxprediction}{\SI{2.6}{\micro \gram \per \day}\xspace}
\newcommand{\PMotwoflux}{\SI{4.2}{\micro \gram \per \day}\xspace}
\newcommand{\PBtdistmeasured}{\SI{0.42}{\hour}\xspace}
\newcommand{\PBhmeasured}{\SI{8.3e-5}{\meter \per \second}\xspace}
\newcommand{\lxeflowempty}{\SI{0.32}{\liter \per \minute}\xspace}
\newcommand{\lxeflowQfive}{\SI{0.15}{\liter \per \minute}\xspace}
\newcommand{\lxeflowAPI}{\SI{0.2}{\liter \per \minute}\xspace}
\newcommand{\epsilonQfive}{\ensuremath{0.92}\xspace}
\newcommand{\epsilonAPI}{\ensuremath{0.66}\xspace}
\newcommand{\apimass}{\SI{300}{\gram}\xspace}
\newcommand{\kotwoval}{\SI[per-mode=power]{1.77e11}{\per \molarity \per \second}\xspace}
\newcommand{\driftfieldval}{\SI{140}{\volt \per \centi \meter}\xspace}
\newcommand{\pmpressureval}{\SI{2}{\bara}\xspace}
\newcommand{\lxedensityval}{\SI{2.88}{\kilo \gram \per \liter}\xspace}
\newcommand{\Tdistval}{\SI{0.42}{\hour}\xspace}
\newcommand{\fval}{\SI{0.937}{}\xspace}
\newcommand{\henryratioval}{\SI{2.44}{}\xspace}
\newcommand{\diffusivityval}{\SI{5.8e-6}{\square \meter \per \second}\xspace}

\usepackage[switch]{lineno}
\newcommand{\sirange}[3]{\SIrange[range-phrase=--, range-units=single]{#1}{#2}{#3}}
\newcommand{\comment}[1]{}

\begin{document}

\title{Liquid-phase purification for multi-tonne xenon detectors}

\author{G.~Plante\thanksref{addr0}
\and
E.~Aprile\thanksref{addr0}
\and
J.~Howlett\thanksref{addr0,email3}
\and
Y.~Zhang\thanksref{addr0}
}
\thankstext{email3}{\email{joseph.howlett@columbia.edu}}

%\author{G. Plante\inst{1} \and E. Aprile\inst{1} \and %J. Howlett\inst{1}\footnote{\email{joseph.howlett@columbia.edu}} \and Y. Zhang\inst{1}
%}                     % Do not remove
%

%\author*[1]{\fnm{Guillaume} \sur{Plante}}\email{iauthor@gmail.com}

%\affil*[1]{\orgdiv{Physics Department}, \orgname{Columbia University}, \orgaddress{\city{New York}, \postcode{10027}, \state{New York}, \country{USA}}}

\institute{Physics Department, Columbia University, New York, 10027 New York, USA \label{addr0}}

%\institute{Physics Department, Columbia University, New York, 10027 New York, USA}

%

\maketitle
\flushbottom

\begin{abstract}
As liquid xenon detectors grow in scale, novel techniques are required to maintain sufficient purity for char\-ges to survive across longer drift paths. The Xeclipse facility at Columbia University was built to test the removal of electronegative impurities through cryogenic filtration powered by a liquid xenon pump, enabling a far higher mass flow rate than gas-phase purification through heated getters. In this paper, we present results from Xeclipse, including measured oxygen removal rates for two sorbent materials, which were used to guide the design and commissioning of the XENONnT liquid purification system. Thanks to this innovation, XENONnT has achieved an electron lifetime greater than \SI{10}{\milli \second} in an $\sim \SI{8.6}{\tonne}$ total mass, perhaps the highest purity ever measured in a liquid xenon detector.
\end{abstract}

\section{Introduction}
\label{sec:intro}

%\subsection{Purification in Liquid Xenon Dark Matter Searches}

% super basic LXe dm intro
The field of dark matter (DM) direct detection has been led for many years by two-phase time-projection chambers\\ (TPCs) using liquid xenon (LXe) as a target and detection medium~\cite{xenon1t_tonyear,lux,pandax4t}. 
Within the XENON project in particular, LXeTPCs with increasing target mass and drift length but diminishing background rate were developed to search for the keV-scale nuclear recoils (NRs) produced by hypothetical Weakly Interacting Massive Particles (WIMPs), one of the leading particle DM candidates~\cite{roszkowski}. 
The current experiment, XENONnT~\cite{xenonnt}, uses a LXeTPC with a drift length of \SI{1.5}{\meter} inside a cryostat filled with $\sim \SI{8.6}{\tonne}$ of ultra-pure LXe.
Two-phase LXeTPCs rely on the efficient collection of ionization electrons produced by recoiling xenon nuclei. 
This signal is generated by drifting electrons upward through the LXe toward an anode, by applying a uniform electric field.
As the electrons drift upward, some are captured by electronegative impurities, reducing the signal. The fraction captured over a given drift increases with the concentration of impurities and their electron attachment rate constant~\cite{bakale}.

%The fidelity of the S2 is limited by the purity of the LXe. As the ionized electrons drift upward, some are captured on electronegative impurities, reducing the consequent S2. The fraction captured over a given drift is proportional to the concentrations of impurities and their rate constants for electron attachment~\cite{bakale}.
% effect on sensitivity
If a substantial number of electrons are lost, the signals are no longer directly proportional to the energy deposited, requiring correction based on the depth of the interaction. In particular, signals from $\sim \SI{1}{\kilo \electronvolt}$ nuclear recoils can fall below the energy threshold, limiting the sensitivity to light DM and solar neutrinos~\cite{xenon1t_cevns}. 
The purity is typically expressed via the ``electron lifetime'' $\tau_e$, the time over which the number of drifting electrons $N_e$ will be reduced by a factor $1/e$ due to attachment:
\begin{equation} \label{eq:elifetime}
N_e(t) = N_e(0)\ e^{-t/\tau_e} = N_e(0)\ e^{-\frac{z}{v_d \tau_e}}.
\end{equation}
Here, $z$ is the length of the drift and $v_d$ is the drift velocity. The electron lifetime is related to the concentrations of impurities and their rate constants of attachment through
\begin{equation} \label{eq:rateconstant}
\tau_e = \frac{1}{\sum k_i C_i} = \frac{1}{k_{\otwo} C_{\otwo}},
\end{equation}
where the \otwo-equivalent concentration $C_{\otwo}$ is often used as the benchmark of purity since \otwo is usually the dominant contributor to charge signal attenuation in LXeTPCs~\cite{bakale}.
For the entirety of this work, the \otwo-equivalent mole fraction $x_{\otwo}$ (often expressed in parts per billion or ppb) is used interchangeably with electron lifetime, according to the relation

\begin{equation}
    \frac{1}{\tau_e} = k_{\otwo} C_{\otwo} = k_{\otwo} x_{\otwo} \frac{\rho_\mathrm{LXe}} {\mmxe} = \frac{x_{\otwo}}{\SI{257}{\micro \second \cdot ppb}}.
\end{equation}
Here, the LXe density $\rho_\mathrm{LXe}=\lxedensityval$ is taken at the approximate operating pressure of \pmpressureval~\cite{nist}, and $\mmxe=\SI{131}{\gram \per \mole}$ is the molar mass of xenon.
The rate constant $k_{\otwo}=\kotwoval$ is taken from Fig.~2 in~\cite{bakale} for the $\driftfieldval$ drift field used throughout this work.

% purification in existing expts
Electronegative impurities in the LXe increase over time due to continuous desorption from detector materials. 
Maintaining components under vacuum prior to filling the cryostat with xenon can reduce the subsequent rate of desorption, but to achieve a sufficient electron lifetime, continuous removal of impurities is required to compensate for their ingress.
In all LXe experiments prior to XENONnT, the removal of electronegatives such as \otwo to preserve ionization signals has been achieved by passing gaseous xenon (GXe) through a gas purifier containing a high temperature ($\approx \SI{400}{\celsius}$) zirconium alloy which absorbes impurities~\cite{saes}. The xenon must first be evaporated, and subsequently re-condensed.
This is typically performed with minimal heat input through the use of highly efficient LXe-GXe heat exchangers, and with flow driven by GXe pumps~\cite{pandax4t,xenon1t_tonyear,lz_tdr}.

\begin{figure}[t]
\centering
\includegraphics[width=0.5\textwidth]{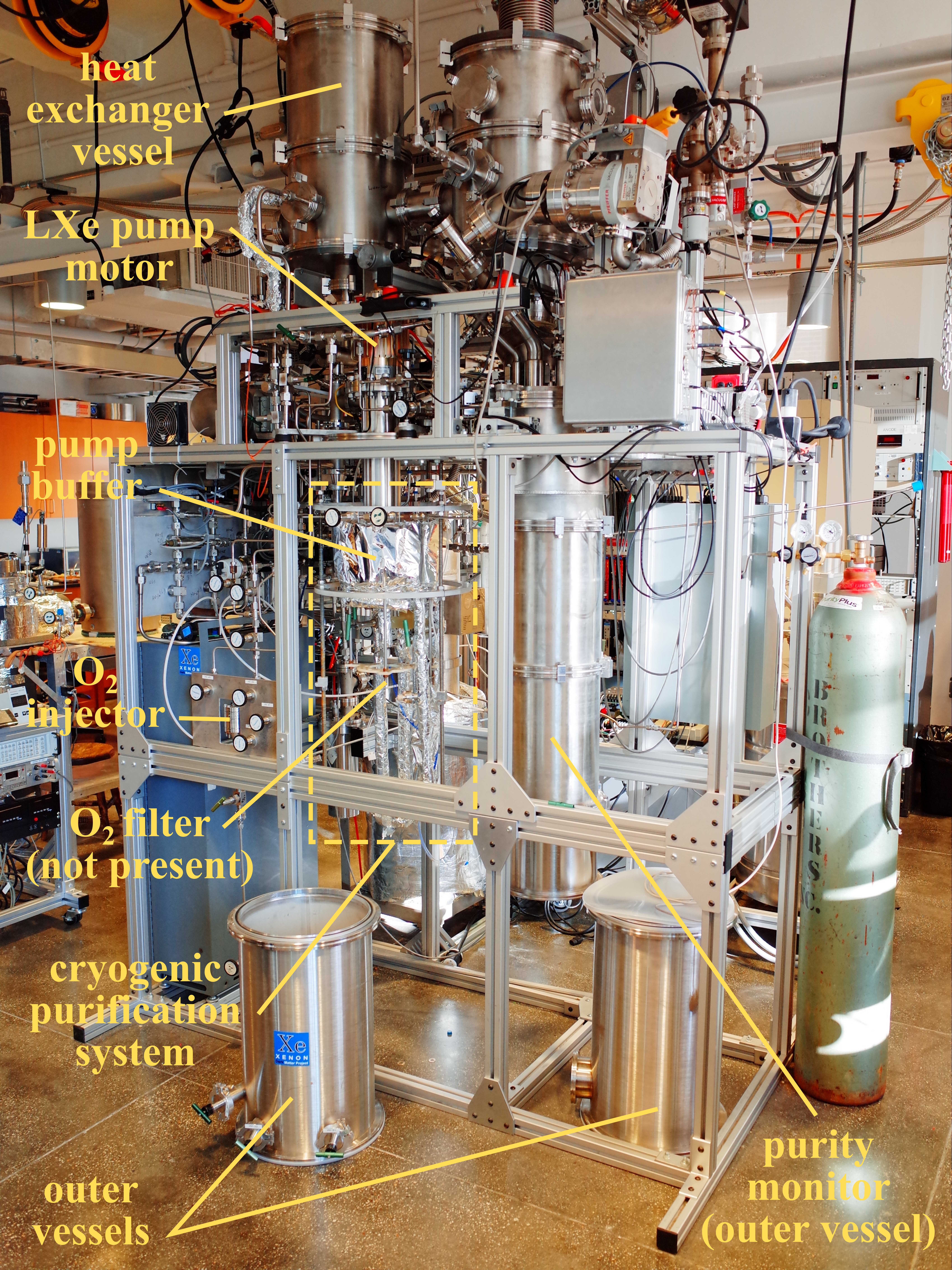}
\caption{The Xeclipse apparatus. The closed vessel of the purity monitor is shown on the right, and the exposed cryogenic purification system on the left with its outer vessel removed.}
\label{fig:physical}
\end{figure}

%\subsection{Limitations and Challenges}
% difficulty of scale (drift length, materials outgassing)
As TPCs have grown in size to become more sensitive to DM, so too has the amount of \otwo-desorbing materials they contain and the time required to reach equilibrium purity, requiring more efficient purification to compensate.
% liquid vs gas
Improvements have heretofore been achieved through the use of more powerful GXe pumps and the combination of multiple pumps in parallel~\cite{xenon1t_instrumentation}, but the pressure limitations of existing pumps, increased heat input, and difficulty maintaining the required pressure and liquid level stability of the TPC make further scaling of the current paradigm a challenge. 
The use of cryogenic liquid pump technology adapted for LXe has the potential to greatly improve the mass flow, due to the factor $\sim 300$ higher density of the cryogenic liquid compared to the room temperature gas at a typical pressure of \pmpressureval. 
To make this paradigm-shift in LXe purification possible, however, a method that cryogenically filters electronegative impurities with high efficiency is required, while meeting the strict radio-purity requirements of direct detection. 
Circulation and filtration in the liquid phase has been established as a standard purification method in liquid argon TPC (LArTPC) experiments aiming to study beam and reactor neutrinos~\cite{icarus,larpur,duneCDR,argoneut}. 
The sorbent beds used by these experiments for \otwo removal characteristically contain sorbent pellets with a high-surface-area copper face~\cite{q5,microboone,argoneut}. As LAr is passed through the filters, impurities react with and become permanently bound to the copper surface. 
The high surface-to-volume ratio of the filter creates a high reactivity, and consequently filtration efficiency, for a relatively small sorbent mass. 
Furthermore, the oxide buildup on the filter, which eventually begins to deplete its reactivity as pure copper sites become scarcely available, can be straightforwardly removed using a purge gas containing a small (\SI{5}{\percent}) concentration of hydrogen, which induces reduction reactions at high ($\approx\SI{200}{\celsius}$) temperature~\cite{q5}. 
While these filters pose no great threat of background introduction for LArTPC neutrino experiments, the low-background nature of DM direct detection experiments presents a unique challenge. The dominant background in LXeTPC DM searches comes from the $\beta$-decay of $^{214}$Pb, a daughter of \rn, which is part of the $^{238}$U decay-chain and emanates continuously from detector materials.
The emanation of \rn tends to increase with the surface area of materials used, and thus poses a challenge when weighed against the high surface areas required for efficient filtration of electronegative impurities.

\begin{figure*}[ht]
\includegraphics[width=1.\textwidth]{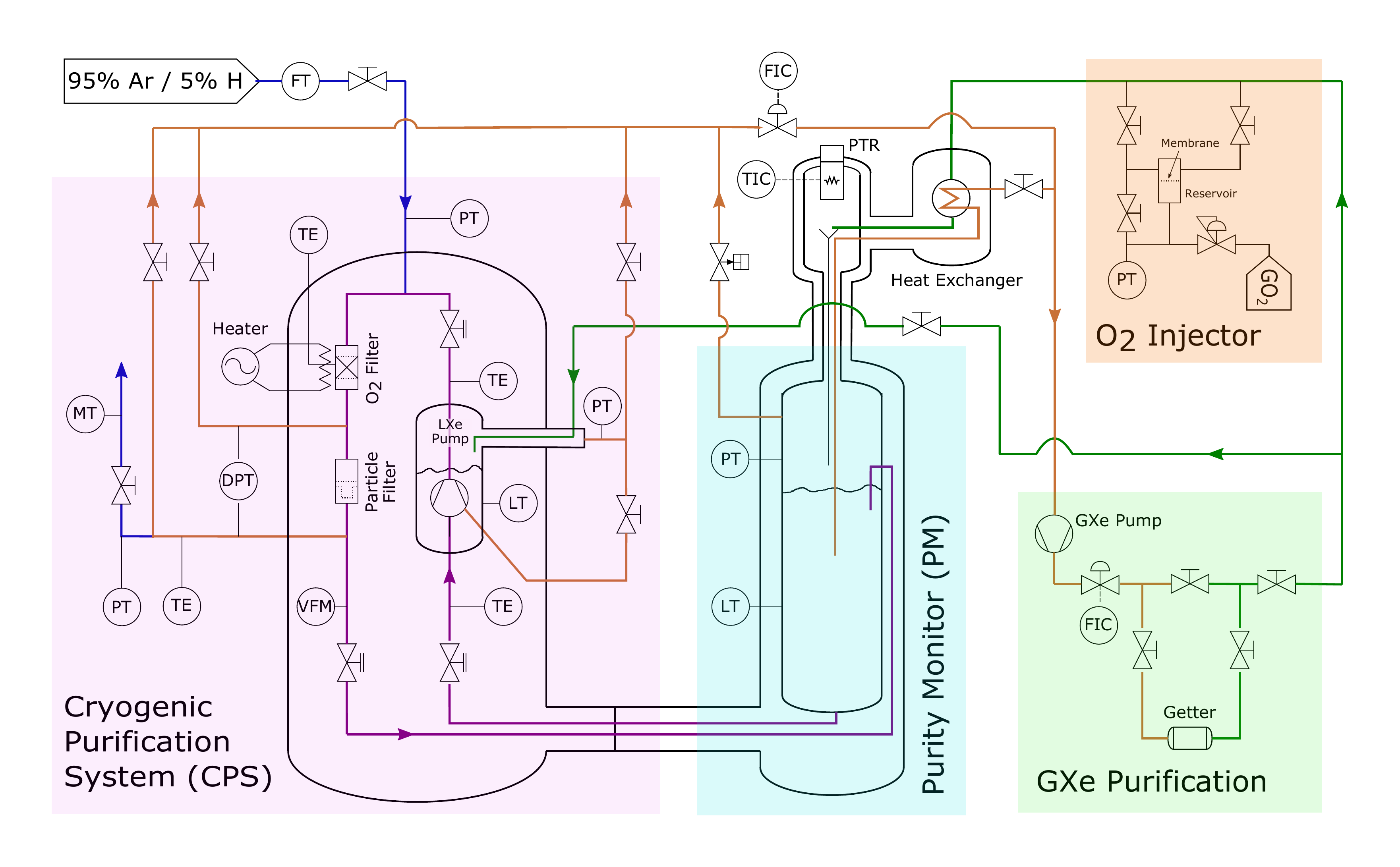}
\caption{Simplified process and instrumentation diagram (P\&ID) of Xeclipse, with the constituent subsystems delineated by boxes. 
Two independent vacuum-insulated stainless steel cryostats house the Cryogenic Purification System (CPS) and the Purity Monitor (PM). 
Also shown are the exterior gas handling systems, including the GXe Purification loop using traditional heated getter purification, and the \otwo injector used to control the rate at which impurities are introduced into the PM.
The xenon flow lines are grouped by color to illustrate the typical configuration. Orange-brown lines carry GXe to the getter, and green lines carry purified GXe from the getter outlet to the PM through the heat exchanger and to the LXe pump buffer GXe volume.
The vacuum-insulated LXe circulation lines are shown in purple, and the filter regeneration lines are shown in blue. Also shown are key pressure transducers (PT), temperature elements (TE) and controller (TIC), liquid level transducers (LT), flow transducers (FT) and controllers (FIC), moisture transducer (MT), the differential pressure transducer used to measure the LXe flow (DPT), and the void fraction meter (VFM) described in the text.}
\label{fig:pid}
\end{figure*} 

%\subsection{Liquid-Phase Xenon Purification}
The Xenon Cryogenic Liquid Purification Setup (Xeclipse) was designed to address the purification challenge posed by the transition from ton-scale LXeTPCs to the multi-tonne scale within the XENON project by testing the efficacy of xenon purification in the liquid phase. 
The apparatus was designed to continuously measure the purity of a LXe volume while recirculating the LXe through both a traditional gas-phase purification loop and a parallel liquid-phase loop. By testing a cryogenic filter while simultaneously injecting \otwo at a fixed, known rate, the apparatus can evaluate the rate of \otwo removed by the filter.
This paper describes the system and its results for two sorbent materials tested for use in XENONnT~\cite{xenonnt}. 
Sec.~\ref{sec:setup} describes the setup and its components.
In Sec.~\ref{sec:model}, a model of \otwo transport between the system's sub-volumes is developed which makes it possible to disentangle the dominant systematics and infer the rate of \otwo removal by each filter.
In Sec.~\ref{sec:results}, the efficiencies of the filters are evaluated in view of their \rn emanation rates, and their viability as purification methods in multi-tonne detectors is discussed.

% Introduction to Xeclipse
\section{Experimental Apparatus}
\label{sec:setup}

% Simplified Xeclipse P&ID
% Maybe some more schematic representation, with clear divisions of the "subsystems": CPS, PM, cryogenics/emergency, GXePUR
%\subsection{Xenon Service and Cooling}
% decided to just make this introductory
% Xeclipse utilizes the demonstrator infrastructure with modifications to test circulation/filtering of LXe.
Xeclipse, shown in Fig.~\ref{fig:physical} and represented schematically in Fig.~\ref{fig:pid}, is composed of two major subsystems: one to continuously measure the LXe's purity, and one to recirculate and purify it. 
The first is a Purity Monitor (PM)~\cite{icarus_pm}, contained in the existing vacuum-insulated vessel which previously housed the XENON1T Demonstrator~\cite{demo}.
The PM is connected by an inlet and outlet line to the cryogenic purification system (CPS), which is housed in a separate vacuum-insulated vessel.
% Two-Vessel CAD showing outer vessel separation
The two insulation volumes are entirely separated by a wall enclosing the inlet and outlet lines. This makes it possible to open the outer vessel containing the CPS and make necessary modifications while the PM is stably full of LXe.

Xeclipse utilizes the existing infrastructure for handling and cooling the required $\sim 50$ kg of LXe from the XENON1T Demonstrator.
% Cryogenics system and circulation using QDrive through hot getter same as [cite demo paper]
The xenon is cooled by a pulse tube refrigerator (PTR) which delivers 200 W of cooling power at \newline \SI{165}{\kelvin}, as detailed in \cite{demo}. 
% LN2 HE?
The GXe purification apparatus described therein is also utilized, with the capability of drawing xenon from the liquid and gas phase in any combination, warming through a heat exchanger, passing through a SAES Monotorr heated getter gas purifier (model PS4MT50R1), and subsequently re-condensing the xenon in the heat exchanger.
The circulation flow is driven by a gas pump with magnetically suspended pistons driven by linear motors (Chart, model 2S132C-X~\cite{qdrive}), and is controlled stably up to $\sim 45$ standard liters per minute (SLPM) through a gas mass flow controller (Teledyne-Hastings, model HFC-303). 
This ``getter purification'' apparatus makes it possible to test novel filtration schemes starting from conditions with measurable purity, as well as to directly compare with the traditional evaporated-LXe purification approach.

\subsection{Purity Monitor}
High-precision continuous measurement of ppb-level \otwo concentrations is critical to evaluate the effectiveness of a given xenon purification method. 
The PM is a proven and reliable method of measuring the purity with a high frequency and low measurement uncertainty~\cite{puritymonitor1,puritymonitor2,icarus_pm}. 
The basic principle is to release a large cloud of electrons and drift the cloud a fixed distance through a uniform electric field. 
The size of the cloud is measured at the beginning and end of the drift using the current induced on a cathode and anode as it moves away from and toward them, respectively. 
Each of these two electrodes is equipped with a grid to shield it from the effects of the cloud except when it is drifting in the space between the electrode and its grid. 
The electron lifetime can be directly calculated from the ratio of the two induced currents and their time-separation.

\begin{figure}[th]
\centering
\includegraphics[width=0.8\columnwidth]{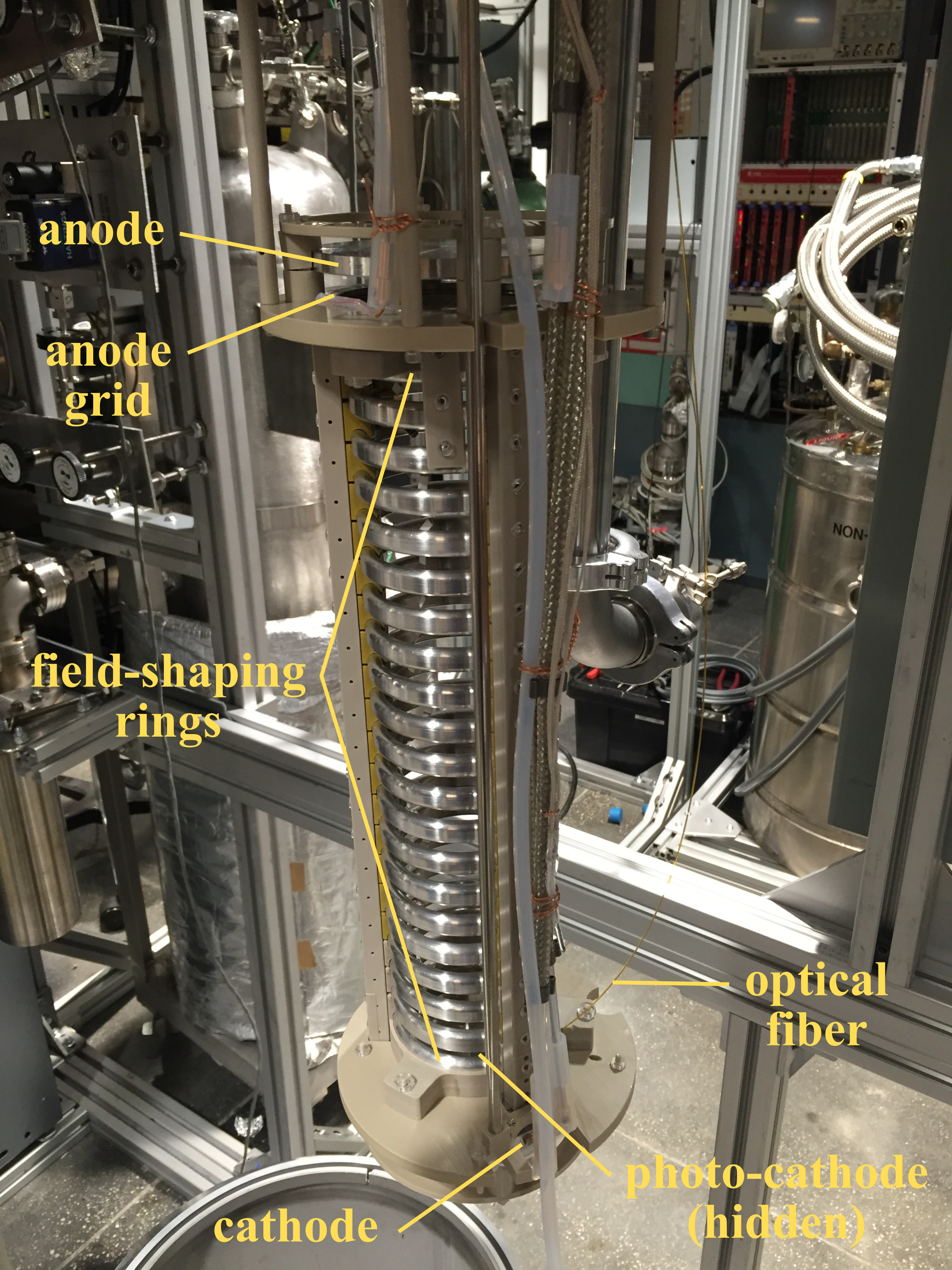}
\caption{The Xeclipse Purity Monitor (PM). The field-shaping rings from the XENON1T Demonstrator~\cite{demo} surround the parallel-grid ionization chamber. The thin, gold-colored optical fiber is shown along the right side of the PM, curving in to face the photo-cathode below.}
\label{fig:pm_photo}
\end{figure}

% We installed a PM inside the existing LXe vessel from demonstrator. Specs of drift length \& electrodes.
The Xeclipse PM, shown in Fig.~\ref{fig:pm_photo}, utilizes the field cage, cooling systems, and heat exchanger from the XENON1T Demonstrator~\cite{demo}. 
A parallel grid ionization chamber is installed \SI{22.7}{\milli \meter} above a photocathode where the electron cloud is released (biased to \SI{-4990}{\V}).
The chamber comprises a \SI{35.1}{\centi \meter} drift length between the cathode grid (biased to \SI{-4840}{\V}) and anode grid (biased to \SI{0}{\V}), yielding a drift field of \driftfieldval. 
A column of field-shaping rings, with an inner diameter of \SI{64}{\milli \meter}, encloses the entire drift length, and an anode is placed \SI{12.8}{\milli \meter} above the anode grid to collect the electrons.
Two charge-sensitive pre-amplifiers (Amptek A250) read out and integrate the induced current on the photo-cathode during the initial drift toward the cathode grid, and the induced current on the anode (biased to \SI{360}{\V}) during the final drift from the anode grid. 
The field more than doubles as the cloud passes each grid, ensuring that nearly 100\% of electrons pass through both~\cite{gridtheory}.
A COMSOL simulation of the Xeclipse field cage and two flat electrodes confirmed that the field lines within the diameter of the impinging light all terminate at the anode~\cite{comsol}.

%As the electron cloud drifts from the photo-cathode to the cathode grid, a current is induced on the cathode whose integral is proportional to the size of the initial electron cloud. During the longer drift between the two grids, no current is induced on either the cathode or the anode. As the surviving electrons drift from the anode grid to the anode, they induce a current on the anode. The induced initial current on the cathode and final current on the anode are read out separately by two charge-sensitive pre-amplifiers (Amptek A250) after capacitive decoupling from the HV bias on the electrodes.

% Photocathode, Lamp, fiber, feedthrough specs.
The electron cloud is produced by the photo-electric effect when light impinging on the photo-cathode exceeds its work function.
The photo-cathode was constructed by depositing \SI{50}{\text{\AA}} of titanium and \SI{1000}{\text{\AA}} of gold on a \SI{2.5}{\centi \meter} outer diameter aluminum disc, which is seated in a larger (\SI{4}{''} outer diameter) aluminum disc. 
Light (\sirange{190}{2000}{\nano \meter}) is produced from a Hamamatsu \SI{60}{\watt} xenon flash lamp~\cite{hamamatsu} and directed through a light guide into a quartz optical fiber. The fiber carries the light into the vacuum chamber through a \SI{600}{\micro \meter} core optical vacuum feedthrough, and is positioned so exiting light is directed onto the gold surface. 
The flash lamp releases 1 Joule of energy, liberating $\sim 10^6$~electrons after losses due to the light guide, fiber, and quantum efficiency of the photo-cathode.
% Preamplifier circuit diagram

% Waveform digitization and saving. Pre-amplifier calibration via pulse generator. [Example waveforms and conversion to lifetime here
The output of each of the two pre-amplifiers is shaped by a spectroscopy amplifier and fed to a digitizer which is externally triggered using the signal triggering the xenon flash lamp. The gain on the two amplification chains are calibrated regularly, by checking the output voltages from a pulse generator signal at various amplitudes on each amplifier. The responses were consistently linear within the range \sirange{0.05}{1.0}{\volt} (corresponding to \sirange{0.02}{0.45}{\pico \coulomb}), which covers the integral charge of the electron clouds typically produced at the photo-cathode.

%The PM LXe is recirculated through the heat exchanger and getter as in~\cite{demo} to maintain a measurable purity, with optional circulation of the GXe volume as well. 
%Closure of the isolation valves to the CPS allows operation in ``PM-only'' mode, with all the xenon confined to the PM, to test the impact of getter purification, oxygen injection, and the evolution of natural oxygen sources.

\subsection{Cryogenic Purification System}
The first challenge in designing the Cryogenic Purification System (CPS) was to achieve a stable and sufficiently high flow of LXe. 
This demands careful control of the thermodynamic conditions everywhere along the flow path to avoid saturation (boiling) of the LXe despite large pressure differences introduced by several vertical passes. 
The second challenge was to allow for the installation, replacement, and in-situ regeneration or activation of various filters. 
The components of the CPS are detailed here, following their order along the LXe flow path.

%\subsubsection{LXe Pump}
% LXe pump: Describe the pump and the vessel it was submerged in, as well as TE's, PT's, valves, and levelmeter used to prime pump and avoid cavitation.
The largest component is the LXe pump and its containment vessel. 
The pump is a BNCP-62-000 centrifugal pump from Barber Nichols~\cite{barber_nichols}. 
% buffer
A primary concern for cryogenic liquid pumps is the phenomenon of cavitation: If the operating pressure at the pump is too close to the vapor pressure of the liquid, the rotation of the impeller may reduce the pressure at its inlet such that gas bubbles form and collapse, releasing shock waves that can cause significant wear on the impeller and instabilities in the delivered liquid flow~\cite{dixon}. 
Cavitation can be avoided through careful control of the thermodynamic properties at the impeller.
The impeller is submerged in a containment ``buffer'' vessel \SI{27}{\centi \meter} in diameter and \SI{7}{\centi \meter} in height.
Pressure and temperature transducers measure the pressure in the buffer and the temperature of its inlet tube, and a \SI{40}{\milli \meter} capacitive levelmeter measures the liquid level to a precision of \SI{1.3}{\milli \meter}.
The pump was controlled using a Toshiba VF-S11 inverter with a typical drive frequency of \SI{23}{\hertz}. 

% starting pump
Starting the LXe pump requires submerging the impeller in liquid, a process called ``priming''~\cite{handbook}.
In Xeclipse, this is achieved by opening a valve that connects the interior volume of the pump to the GXe volume above the PM, reducing the pressure such that the LXe is allowed to rise and submerge the impeller.
The inverter is then used to control the impeller rotation frequency and achieve the desired pressure difference \deltaP.
The inverter was programmed to shut off the pump if $\deltaP$ falls below \SI{200}{\milli \bar} for more than \SI{40}{\second}, since this indicates that the pump volume has become filled with gas, or ``gas-bound'', preventing LXe flow.

It was discovered that the LXe pump contains a significant source of \otwo-equivalent impurities which is continually introduced into the xenon. 
Since the pump is mechanically sealed rather than magnetically coupled, the rotating shaft passes from the electric motor directly into the pump casing containing the impeller.
As the pump is running, the GXe above the pump inlet is therefore in contact with the motor, which is close to room temperature.
A consequently high concentration of impurities like \otwo in this GXe volume can lead to transfer into the LXe at the liquid-gas interface.
To mitigate this impurity source, GXe is continuously extracted from the pump buffer and LXe pump itself and fed to the getter at the port shown in Fig.~\ref{fig:pid}. Clean GXe from the getter outlet is returned to the pump buffer but not to the pump interior volume. 
This may result in a low \otwo concentration in the pump buffer GXe, but a relatively high concentration in the pump interior. 
The observed impact on the \otwo exchange at the liquid-gas interfaces is detailed in Sec.~\ref{ssec:gas_exchange}.

%Due to the proximity of the operating pressure at the pump to the LXe saturation pressure, care must be taken to avoid cavitation and gas binding of the pump.
At a few points in the CPS circuit, such as vertical U-bends, small static gas volumes have a tendency to form and grow due to heat input (and low fluid velocity), eventually causing flow instabilities or loss of \deltaP (gas binding).
Metering valves were installed at these points, connecting the gas volumes to the GXe purification loop inlet to prevent them from growing and thus maintain stable liquid flow. 
%Extremely fine tuning of the valves was required to avoid cavitation. 
%The valve openings were optimized to a flow coefficient ($C_V$) of $\approx~0.01$ based on the VFM reading to minimize bubbles of GXe in the LXe flow \cite{metervalve}.

%\subsubsection{\otwo Filter}
%\label{sssec:otwofilter}
% \otwo filter vessel: A modular filter vessel which can be removed and modified. Typically composed of a nipple or tube with sintered disc filters on either end to hold the filter material being tested inside. Fixed to G-10 frame for easy installation/removal
The LXe purification occurs within the filter vessel. The CPS is constructed such that this vessel can be substituted, heated, pumped, or otherwise modified without disassembling any other part of the system. 
Its structure and dimensions depend on the sorbent material being tested, but it is typically composed of a stainless steel tube containing a packed bed of sorbent pellets, encased by stainless steel sintered discs at the inlet and outlet. 
%Substitution of the filter vessel is allowed from below when the bottom flange of the CPS outer vessel is removed. 
%Vertical rods attached to the assembly from above are guided into through-holes in a G-10 frame fixed to the bottom flange of the pump buffer. Nuts must then be inserted at the height of the flange connecting the two outer vessel nipples, without removing the lower nipple, since it is connected to the CPS isolation valves. Vertical rods connect the lower nipple to the top flange of the outer vessel, supporting it while it is lowered about one inch to insert the nuts and fix the filter vessel to the G-10 frame, and subsequently to close the \sfrac{1}{2}'' VCR fitting at the filter outlet. The \sfrac{1}{2}'' VCR fitting at the inlet is closed from below.

% Main candidate of \otwo filter is Q5, works with pure Cu
Two sorbent materials were tested in this work. The first is Engelhard Q5 copper-impregnated spheres (equivalent to BASF ``CU-0226 S'')~\cite{basf}, an oxygen getter catalyst composed of copper oxide (13\% by weight) deposited on an alumina carrier in 14x28 mm mesh beads with a high (200 m$^2$/g) surface area. The beads are activated/regenerated by reduction using a gas mixture of \SI{95}{\percent} argon and \SI{5}{\percent} hydrogen around \SI{200}{\celsius}, which leaves a pure copper surface that can readily react with oxygen. 
The radio-purity of the catalyst was measured to evaluate the tolerable filter mass for multi-tonne LXeTPCs~\cite{xenonnt_radiopurity}.  
The contamination in \ra, the parent nucleus of \rn, was found to be \SI{150}{\milli \becquerel} per kilogram of the catalyst. 
Measurements also showed that about 1/3 of the \rn produced actually emanates from the sorbent material, about \SI[allow-number-unit-breaks]{50}{\milli \becquerel \per \kilo \gram}.
Later tests of a different sample yielded $\sim 7 \times$ higher values for both \ra contamination and \rn emanation~\cite{xenonnt_radiopurity}.

% second candidate NEG St 707
The second is a Non-Evaporable Getter (NEG) formed into small cylindrical pills by compressing a getter alloy powder composed of zirconium, vanadium, and iron. 
This is the same material used inside the high temperature GXe getters described in Sec.~\ref{sec:intro}, whose $^{222}$Rn emanation rate was measured to be \sirange{0.2}{1.2}{\milli \becquerel} for $\sim$\SI{4}{\kilo \gram} of sorbent material in~\cite{xenon1t_radiopurity}. 
At such high  ($\approx \SI{400}{\celsius}$) temperatures, oxygen and other impurities adsorbed on the surface of the pill quickly diffuse into the bulk, leaving the pill surface relatively free of impurities. 
At LXe temperature, we expect the inward diffusion to be much slower, such that impurities accumulate on the surface, eventually limiting the sorption rate. 
The NEG has a much lower rate of \rn emanation, but the total reactive surface area per unit mass is much lower than the Q5 catalyst~\cite{xenonnt_radiopurity}.

% particulate filter
To prevent particulates of sorbent media that escape the filter vessel outlet through the sintered disc from entering the PM, an additional, finer grade sintered tube particulate filter is used (Mott, 2300 series, \SI{0.2}{\micro \meter} media grade). A differential pressure transducer measures the pressure difference across the particulate filter, which is used to calculate the LXe flow in the circuit. 
The particulate filter permeability was calibrated using a controlled flow of GXe between~\SIrange[range-phrase=\ and\ , range-units=single]{1}{4}{\slpm}, and was found to be \SI{37.0}{SCCM \per \milli \bar}\footnote{SCCM: standard cubic centimeters per minute}, roughly half of the theoretical value for the sintered media grade and dimension. The ratio of theoretical LXe and GXe permeability was used to convert this to \SI[per-mode=reciprocal]{1.54}{\liter \per \minute \per \bar} for LXe.

To enable evacuation and regeneration of the filter vessel containing the Q5 catalyst in-situ, the CPS includes lines to pass the argon-hydrogen mixture gas through the vessel and out to air. 
Resistive heaters attached to the outside of the vessel are used in a PID loop with a thermocouple to control the temperature, and a moisture transducer at the outlet to air measures the water content, in order to monitor the progress of the reduction reaction.

%\begin{figure}[ht]
%\includegraphics[width=1.\columnwidth]{figures/xeclipse_unrolled_placeholder.png}
%\caption{Schematic representation of Xeclipse, with the internal tubing of the CPS ``un-rolled'' to make the flow path clear. Also shown are locations of pressure transducers (``PT'') and temperature elements (``TE''), and the locations of gas outlets to the getter. }
%\label{fig:unrolled}
%\end{figure}

%\subsubsection{Void Fraction Meter}
% Void fraction meter
To monitor liquid flow stability and watch for cavitation at the liquid pump, an instrument was designed to measure saturation of the pump outlet LXe.
The amount of GXe bubbles contained in this liquid, expressed as the ``void fraction'', is continuously measured using a capacitive void fraction meter (VFM). 
The VFM is a capacitor consisting of several concentric stainless steel cylinders aligned with the flow path. As the LXe flows through the space between the cylinders, any GXe bubbles inside reduce the dielectric permittivity and consequently the capacitance. 
The resulting total capacitance was calculated by simulation, and then its linear dependence on the void fraction was confirmed with calibrations using sub-cooled liquid and pure gas xenon flows.
The VFM location is shown in Fig.~\ref{fig:pid}.

%\subsubsection{Connections to PM}
% Isolation valves for LXe transferral 
%The CPS was designed to allow for online modification or substitution of the filter vessel. 
Two isolation valves at the inlet and outlet of the CPS circuit and one before the filter vessel inlet control the transferal of LXe between the CPS and PM volumes. 
The valves are shown in Fig.~\ref{fig:pid} along the purple LXe circulation path.
The entire LXe mass can be transferred to the PM by halting the liquid pump and closing the valves connecting the gas volumes of the CPS and the PM, causing the CPS pressure to increase due to LXe evaporation, and the LXe to retreat to the PM vessel, until only GXe remains in the CPS.
Subsequently, the isolation valves can be closed and the CPS insulation vessel can be opened without substantial heat input to the PM, and modification or replacement of the filter vessel can be performed without disrupting the PM operation.
Closure of the isolation valves also allows operation in ``PM-only'' mode, with all the xenon confined to the PM, to test the impact of getter purification, oxygen injection, and the evolution of natural oxygen sources.

\subsection{Oxygen Injector}
\label{ssec:o2inj}
%The purity of the LXe at a given time after purification commences is a function of the initial purity, the purification speed and efficiency, and the rate at which new impurities are introduced into the liquid due to sources such as desorption and un-purified gas volumes.
%These sources are difficult to predict, and themselves can change with time and the purity of the LXe, posing a challenge to evaluating the purification efficiency in a controlled way.
An apparatus was designed to inject \otwo into the xenon from a gas bottle at a fixed, known rate.
% schematic
The apparatus consists of a ``reservoir volume'' containing an \otwo-GXe mixture separated from the GXe-purification loop by a permeable silicone rubber membrane with \SI{9}{\milli \meter} diameter and \SI{1.6}{\milli \meter} thickness. 
It can be isolated from or included in the GXe-purification loop after the GXe is purified by the getter (see Fig.~\ref{fig:pid}).
As purified GXe flows past the membrane, \otwo migrates from the reservoir volume at a rate linear in the difference of \otwo partial pressures on either side, entering the returning gas before it is re-condensed in the heat exchanger.
% permeability measurement
The permeability of the membrane was calibrated by filling the reservoir volume with an equal (\SI{1.8}{\bar}) pressure of GXe to purification loop side without any flow, then adding \SI{900}{mbar} of \otwo to the reservoir volume, and monitoring the decrease in reservoir pressure over three days. 
Fitting a linear trend to the decrease gave an \otwo permeation rate (per unit partial pressure in the reservoir volume) of \newline \SI[per-mode=power]{3.082 \pm 0.023 e-11}{\gram \per \second \per \milli \bar}. 
This measurement \newline matched well with a coarser estimate using a residual gas analyzer to monitor the \otwo peak in the purification loop volume under vacuum, with a nitrogen-oxygen mixture in the reservoir volume.

%For some tests, such as evaluating the maximum \otwo mass that a given filter can adsorb before saturation begins to occur, it is advantageous to inject a larger known mass of \otwo promptly. This ``prompt injection'' is performed by combining GXe and \otwo on the upstream side of the membrane in a known ratio, then allowing this mixture to expand into a small, known volume such that the resulting \otwo mass inside the volume is on the order of micrograms.

\begin{figure*}[ht]
\includegraphics[width=1.\textwidth]{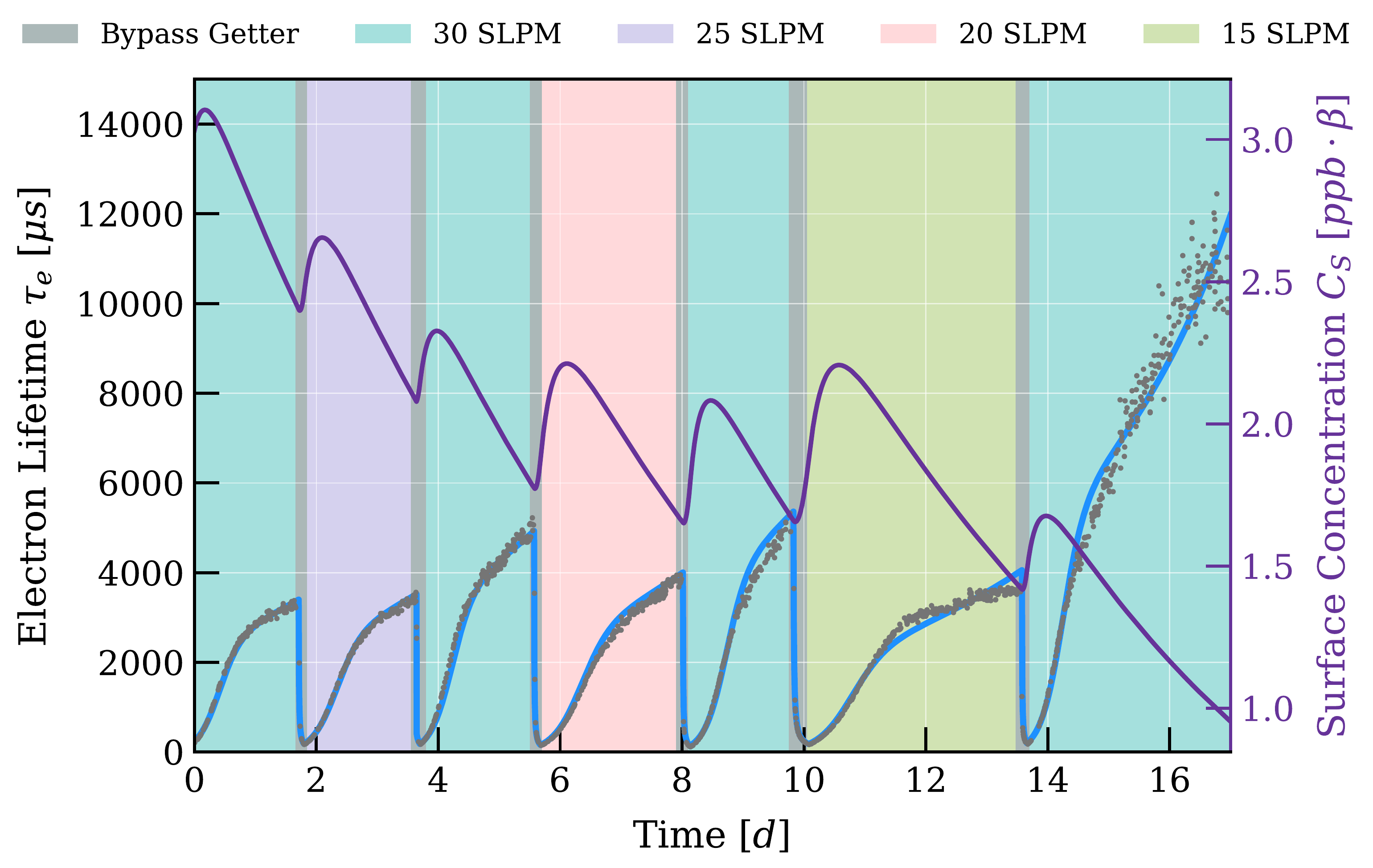}
\caption{PM data from test of traditional evaporated-LXe purification at different mass flows. The data are compared to a model in which impurities are exchanged with a surface in the LXe (Eq.~\ref{eq:wall_model}), with the surface concentration shown in purple.}
\label{fig:getter-model}
\end{figure*} 

\section{Oxygen Transport Model}
\label{sec:model}

The PM measures only the average \otwo concentration along the drift of the electron cloud, but the concentration at other locations in the system is in general not known. Neglecting the small volumes of LXe-carrying tubes along the flow path, the \otwo can migrate between five known volumes: LXe in the PM, GXe in the PM, LXe in the liquid pump and its surrounding buffer, GXe above the pump buffer liquid, and GXe in the pump casing itself. Each of these volumes can have a different \otwo desorption rate, with a different time-dependence, and can exchange \otwo with one another. In order to understand these many correlated, unknown rates, we utilize a modeling scheme with coupled differential equations for the various volumes, where parameters are fixed by data from operations in various simplified flow configurations. 
This section develops the model, starting from a generic case and then applying it to the specific case of PM electron lifetimes in PM-only mode. 
This validates its basic features and fixes a nuisance parameter related to \otwo equilibration.
A model of interphase \otwo exchange is then introduced, and the pieces are put together to build the full model of \otwo transport in Xeclipse.
We then describe the methods employed to disentangle the various nuisance parameters so that the model can be matched to the PM data and infer the filtration efficiency of the materials studied.

\subsection{Generic Model} %\label{sssec:getterpur}
In general, the rate of change in the number of impurities in the LXe equals the rate they enter due to all sources present minus the rate they are removed by purification. 
For a generic system with LXe continuously purified at a mass-flow rate \mdot by a method with efficiency $\epsilon$, and with new \otwo introduced at a rate $\Lambda$ (in moles per unit time, sometimes expressed in \SI{}{\micro \gram \per \day} for easy comparison with desorption rates in DM detectors), the mole fraction \xpm of \otwo in the LXe is described by
\begin{equation} \label{eq:generic_model}
\npm \frac{d \xpm}{dt} = \Lambda - \frac{\npm \epsilon f}{\Tlpm} \xpm,
\end{equation}
where \npm is the amount of LXe in moles, \Tlpm is the time to circulate the full mass of LXe \ml, given by $\Tlpm = \ml / \mdot$, and $f$ is a coefficient relating to how rapidly the purity equilibrates throughout the LXe (a value $f=1$ means that the impurities are homogeneously distributed and equilibrate instantly across the volume when \xpm changes).

When $\xpm(t=0)$ is high, the second term on the right hand side of Eq.~\ref{eq:generic_model}~dominates, giving $\xpm(t) \approx \xpm(0) \cdot e^{-t/\Tlpm}$ (for $f=\epsilon = 1$). After some time $t \gg \Tlpm$, however, the rate of \otwo removal becomes comparable to the sources of new \otwo and \xpm becomes flat at 

\begin{equation} \label{eq:asymptotic}
\xpm(t \gg \Tlpm) = \frac{\Tlpm \Lambda}{\epsilon f \npm} = \frac{\Lambda \mmxe}{\epsilon f \mdot}.
\end{equation}

Often, $\Lambda$ is not constant, but can be well described by the time-dependence
\begin{equation} \label{eq:lambdat}
\Lambda(t) = \frac{\Lambda(0)}{1 + t/T_{\sfrac{1}{2}}},
\end{equation}
with ``half-life'' $T_{\sfrac{1}{2}} \gg \Tlpm$.\footnote{Note that this corresponds to the $t^{-1}$ time dependence that is often observed for desorption from real surfaces (e.g.~\cite{ohanlon} p.66).} Then, expansion to first order in $T_{\sfrac{1}{2}}/\Tlpm$ gives

\begin{equation} \label{eq:asymptotic_time}
x(t \gg \Tlpm) \approx \frac{\Lambda_0 \mmxe}{\epsilon f \mdot} \bigg(1 - \frac{t}{T_{\sfrac{1}{2}}}\bigg).
\end{equation}
\begin{figure*}[t]
\includegraphics[width=1.\textwidth]{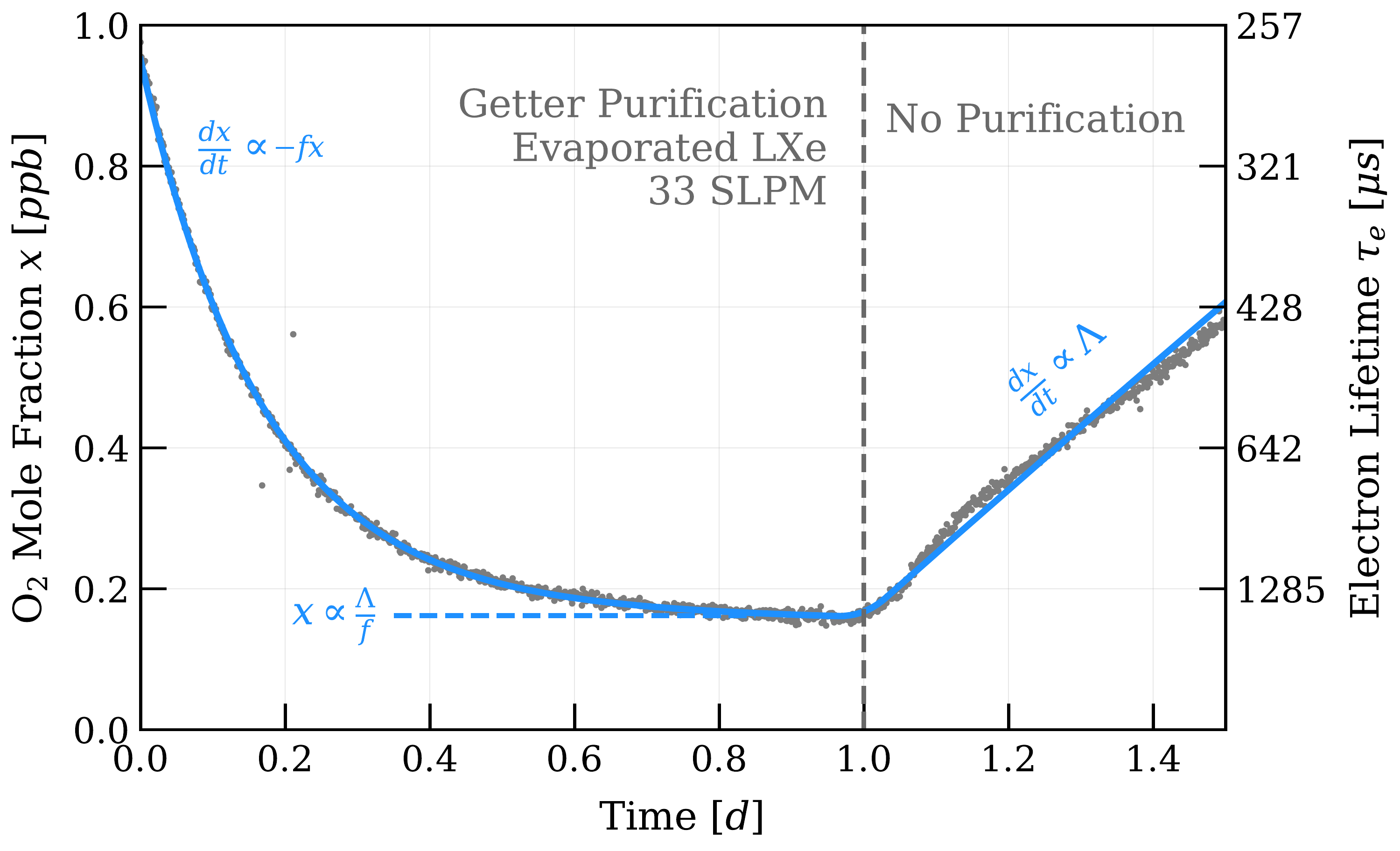}
\caption{Model (blue) of PM \otwo mole fraction (gray) in PM-only mode, before and after stopping circulation through the getter (dashed gray vertical line). The data after the change constrains the \otwo source magnitude $\Lambda$, and the prior data then constrains $f$. The dependence of the dynamics on the two systematics is shown in the initial ``bulk cleaning'' (exponential) phase, the asymptotic ``equilibrium'' phase, and the final (linear) period.}
\label{fig:fcalculation}
\end{figure*} 
The usual sources of \otwo, such as desorption from detector materials, have a complex dependence on time and temperature~\cite{ohanlon}. 
At LXe temperature, the slow decrease in $\Lambda$ during purification can in most cases be modeled as Eq.~\ref{eq:lambdat}, sometimes requiring a small quadratic correction term.
Thus, the typical trend of measured \otwo concentration is an initial exponential decay with time-constant $\Tlpm$, followed by a linear trend defined by the reduction of \otwo sources with time.
When we inject \otwo at a fixed rate large enough to disregard other sources, the time-dependence is removed and we are left with Eq.~\ref{eq:asymptotic}, where $\Lambda \approx \Linj$, the \otwo injection rate.

%\subsubsection{Validating Purification Term}
The basic features of this generic model were tested using Xeclipse in ``PM-only'' mode, with all of the liquid contained in the PM and the isolation valves closed. The liquid is continuously purified through the getter ($\epsilon = 1$~\cite{dobe}) via the HE at several mass flow rates, with no \otwo injection. 
Each iteration was begun from an initial electron lifetime near the PM measurement threshold (\SI{40}{\us}) and lasted until an asymptotic trend with a small, linear time-dependence (Eq.~\ref{eq:lambdat})~was reached. Then, the getter was bypassed in the purification loop to bring the purity quickly back to the initial low value, and the next iteration was begun at a different flow speed.
An iteration at \SI{30}{\slpm} was repeated after each alternate flow rate to observe the time-dependence of $\Lambda$ directly.
The resulting $\Lambda(t)$ calculated from the \SI{30}{\slpm} iterations, however, cannot be straightforwardly interpolated, as it was found to depend on the purity levels during the intervening iterations. Periods of low purity in the liquid have the effect of carrying $\Lambda$ back to an earlier, higher value. This makes sense if, for example, $\Lambda$ is due to a submerged surface which can exchange \otwo with the LXe such that

\begin{equation}  \label{eq:wall_model}
\begin{split}
    \frac{d \xpm}{dt} & = \frac{\Csurf}{\beta \tau_{D,S \rightarrow L}} -  \frac{\xpm}{\tau_{D,L \rightarrow S}} - \frac{f x}{\Tlpm};  \\
    \frac{d \Csurf}{dt} & = \frac{\beta \xpm}{\tau_{D,L \rightarrow S}} - \frac{\Csurf}{\tau_{D,S \rightarrow L}},
\end{split}
\end{equation}
where \Csurf is the concentration on the surface, $\tau_{D,L \rightarrow S}$ and $\tau_{D,S \rightarrow L}$ are the diffusive exchange time-scales from the liquid to surface, and surface to liquid, respectively, and $\beta$ is a coefficient with dimensions that relate the surface and volumetric \otwo concentrations.
Thus, the time-dependence of the rate of \otwo entering the LXe comes from the depletion and accretion of impurities on this surface via exchange with the liquid. Figure~\ref{fig:getter-model}~overlays this model and the electron lifetime data. 
As the getter is bypassed at the end of each iteration and \xpm is restored to a higher value, \Csurf is consequently replenished, but is then reduced to a level which depends on the purification speed of the following iteration.
The success of the model in describing the overall behavior of the time-dependence of the electron lifetime verifies the dependence on \mdot, a basic feature that can be used to describe the initial exponential behavior and asymptotic purity of any system with a fixed purification mass flow.

%\subsubsection{Determining the Coefficient $f$}
%\label{sssec:f}
A measurement in PM-only mode with and without getter purification is used to determine the value of $f$. Purification of the LXe through the getter at \SI{33}{SLPM} was started from an initial purity below the PM measurement range ($\tau_e <\SI{40}{\us}$) and continued until the electron lifetime began to flatten. 
Then, the flow through the getter was stopped, and the electron lifetime was allowed to decrease again due to intrinsic \otwo sources. 
The resulting data is shown in Fig.~\ref{fig:fcalculation} overlayed with the model of Eq.~\ref{eq:generic_model}. 
Here, since the PM contained \SI{45.0}{\kilo \gram} of LXe, $\Tcl=\SI{3.8}{\hour}$ and $\npm=\SI{343.2}{\mole}$. 
The late-time data without purification constrains $\Lambda$, in this case to a constant value of $\SI{3.4e-7}{\mole \per \day}=\SI{10}{\micro \gram \per \day}$, which, when simultaneously considering the earlier trend during purification, leads to an estimate of $f=\fval$.
Physically, a value $f>1$ would indicate that LXe for purification is drawn from a region of higher than average \otwo concentration (e.g. near an \otwo-desorbing surface). The observed value is consistent with nearly homogeneous distribution of impurities throughout the LXe, and relatively rapid mixing of new impurities. 
Since natural convection driven by the total heat transfer from the vertical steel wall induces much higher fluid velocities in the PM than either getter or liquid purification~\cite{bergman}, the impurity mixing coefficient $f$ is taken to be due to convection, and thus constant with respect to circulation speed. We therefore use this value in the full model in Sec.~\ref{ssec:lxepur_model} as well. 
Furthermore, we assume that the homogeneity and mixing rate of impurities in the pump buffer LXe volume is at least as good as in the PM, as it is a smaller volume with a higher heat input per unit surface area. 
This would translate to a value closer to unity. Since this pump buffer mixing coefficient is $<\SI{10}{\percent}$ different from the $f$ measured here, cannot itself be easily measured, and is degenerate with several other model parameters, we use the same value $f=\fval$ in the pump buffer as well.

% Modeling elifetime evolution (PM-only) while turning on/off GXe volume circulation.
\subsection{O$_\text{2}$ Exchange at the Liquid-Gas Interface} \label{ssec:gas_exchange}

When the LXe is purified but the GXe above it is not, equilibrium is disturbed such that \otwo will migrate from the GXe to the LXe, constituting one of the sources of \otwo included in $\Lambda$ in the above cases.
In the CPS pump buffer, however, the GXe volume is continuously purified to mitigate the aforementioned introduction of impurities at the LXe pump. 
Thus, the mole fraction of \otwo in the pump buffer GXe is much less than the LXe due to the smaller mass, resulting in a net positive \otwo exchange from the liquid to the gas volume, which must be considered in the model of LXe purity.
Here, a model based on free convection is developed to describe interphase transport of \otwo in xenon. 
The model is validated using the PM in PM-only mode with continuous purification of the PM GXe. Then, we apply the model to the CPS pump buffer to account for interphase transport in our evaluation of filter efficiencies.

For a given mole fraction of \otwo in LXe \xl, the equilibrium mole fraction of \otwo in the GXe above $x_\mathrm{G,eq}$ for which net zero \otwo transfer takes place at the liquid surface is given by the Henry's law volatility constant:

\begin{equation}
    K_H = \frac{x_\mathrm{G,eq}}{\xl},
\end{equation}
which is calculated to be \henry for \otwo in xenon using the method described in~\cite{pierotti}. 

Using a boundary layer approach~\cite{bergman}, we assume the concentration \Cbl (in moles per unit volume) in the boundary layer just above the liquid surface to be in Henry's-Law equilibrium with the liquid so that

\begin{equation} \label{eq:Ceq}
    \Cbl = \frac{\rhog}{\mmxe} \xg = \frac{\rhog}{\mmxe} K_H \xl.
\end{equation}
The total transport rate of \otwo driven by the difference in concentrations between this boundary layer and the concentration higher in the gas mixture $C_\mathrm{G}$ is then given by

\begin{equation} \label{eq:molar_flux}
    \ndotsurf = \hm \Asurf (\Cbl - C_\mathrm{G}) \approx \hm \Asurf K_H \frac{\rhog}{\mmxe} \xl,
\end{equation}
where \Asurf~is the area of the liquid surface and $\hm$ is the convection mass transfer coefficient (discussed below)~\cite{bergman}. 
Here it is assumed that the \otwo concentration in the bulk of the GXe is much lower than near the surface ($C_\mathrm{G} \ll \Cbl$) due to the rapid purification of the GXe volume. The rate of change of \xpm due to interphase transport \xdotsurf is given by dividing Eq.~\ref{eq:molar_flux} by the amount of LXe, giving
\begin{equation} \label{eq:ppb_flux}
    \xdotsurf = \frac{\hm \Asurf}{\Vl} K_H \frac{\rhog}{\rhol} \xl = \frac{\hm \Asurf}{\henryratioval \Vl} \xl = \frac{\xl}{\Tdist},
\end{equation}
where we define the characteristic time of the mass transfer process at the liquid surface $\Tdist \equiv \frac{\Vl}{\hm \Asurf} \cdot \frac{K_H}{\rhol/\rhog}$, with the substitution $\frac{\rhol/\rhog}{K_H}=2.44$ for saturated xenon at \pmpressureval\footnote{The LXe in the pump buffer is slightly sub-cooled but the change in density is negligible.}, applicable in both the PM and pump buffer.

\begin{figure*}[t]
\includegraphics[width=1.\textwidth]{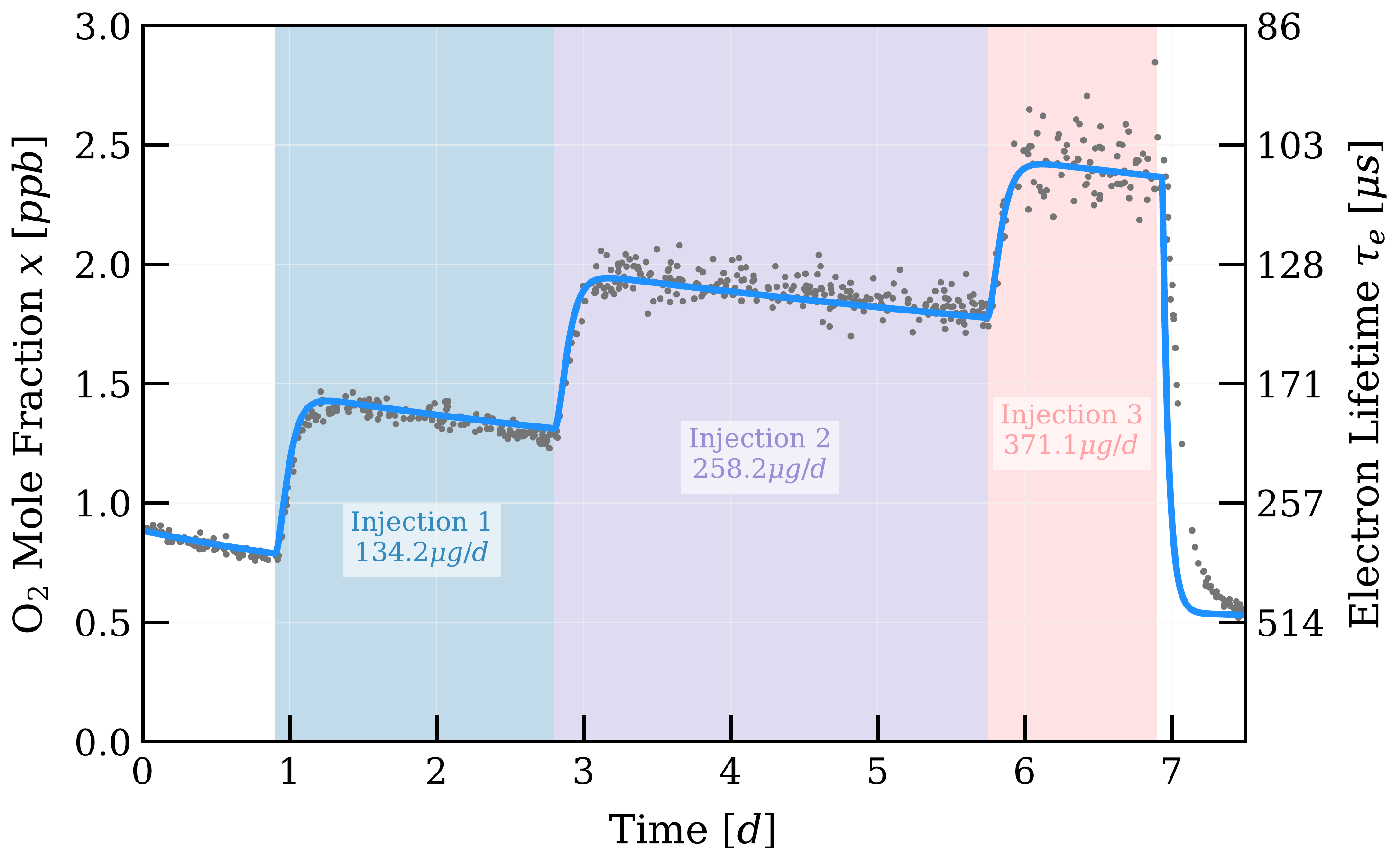}
\caption{Injection procedure performed with an empty tube substituted for the \otwo filter, in order to measure the rate of interphase transport in the LXe pump buffer. The colored regions indicate the three continuous injections. The decrease in PM \otwo mole fraction \xpm during each period is mainly due to the reduction of the dominant \otwo source in the pump buffer volume \Lpb throughout the period shown, though the depletion of \otwo in the finite injector reservoir volume also plays a significant role during the ``Injection 3'' period. 
The LXe flow in the circuit was \lxeflowempty for the full period.}
\label{fig:nofilterfit}
\end{figure*} 

The coefficient $\hm$ is determined from the dimensionless Sherwood number $\mathrm{Sh} = \frac{\hm}{D/L}$, where $D$ is the mass diffusivity of \otwo in the GXe, calculated to be \diffusivityval, and $L$ is the ``characteristic length'' of the convection surface.
For the problem of free convection from a horizontal plane where the gas above the plane contains a lower concentration of the species in question, $L$ is given by the area of the plane divided by its perimeter~\cite{bergman}.
The Sherwood number depends on the dimensionless Rayleigh number $\mathrm{Ra}$, which can be calculated from the thermodynamic properties of the system. The correlation function $\mathrm{Sh}(\mathrm{Ra})$ has been evaluated using a variety of experimental and numerical methods under conditions similar to those in Xeclipse~\cite{lloyd_moran,suriano_yang,mikheyev,goldstein_sparrow,goldstein}. We use $\mathrm{Sh}=0.54\ \mathrm{Ra}^{1/4}$ from \cite{lloyd_moran}.

$\mathrm{Ra}$ is a product of the Schmidt number $\mathrm{Sc}$ and the Grashof number for mass transfer $\mathrm{Gr}_c$.
The Schmidt number is the ratio of the kinematic viscosity of the GXe (viscosity per unit density), calculated to be \SI{7.82e-7}{\square \centi \meter \per \second}, to the mass diffusivity of \otwo, thus $\mathrm{Sc}=0.0616$. The Grashof number for mass transfer is given by (following the assumptions of Eq.~\ref{eq:molar_flux})
\begin{equation}
\label{eq:grashof}
\mathrm{Gr}_c = \frac{g \beta^* (\Cbl - C_\mathrm{G}) L^3}{\nu^2} \approx \frac{g \beta^* L^3}{\nu^2} K_H \frac{\rhog}{\mmxe} \xl,
\end{equation}
where $g$ is the gravitational acceleration, and \\ $\beta^* = -(1/\rhog) (\partial \rhog/ \partial C_{\otwo})$ is the ``specific densification coefficient'' evaluated at the surface, calculated using the CoolProp package (saturated xenon with $\tau_e=\SI{1}{\milli \second}$ and \SI{2}{\bara}, constant over the range of \otwo concentrations measured.)~\cite{coolprop}.

%\subsubsection{Measurement in PM-Only Mode}
To test this model, the interphase \otwo exchange in the PM was measured in PM-only mode and compared to the model's prediction.
The rate of increase of impurities in the LXe with no circulation was subtracted from the rate while purifying the GXe volume above it. 
The difference in these two rates is equal to the rate of \otwo migrating from the LXe to the GXe in the latter case, which was found to be \PMotwoflux.

To make a prediction using the model, we insert the \otwo concentration at the time of this measurement into Eq.~\ref{eq:grashof} to get $\mathrm{Gr}_c$. 
This, along with $L=\PMcharlen$ for the PM liquid surface yields $\mathrm{Sh}=\PMsherwood$ and $\hm=\PMh$, resulting in a predicted \otwo mass flow of \PMotwofluxprediction (using Eq.~\ref{eq:molar_flux}). 
This correspondence within a factor of $2$ between the predicted and observed rates is smaller than the variance between Sherwood numbers in the literature.
%The values of $\mathrm{Sh}$ and $\hm$ resulting for several correlations are shown in Tab.~\ref{table:sherwoods} to demonstrate the variability in the literature.

\begin{figure*}[t]
\includegraphics[width=1.\textwidth]{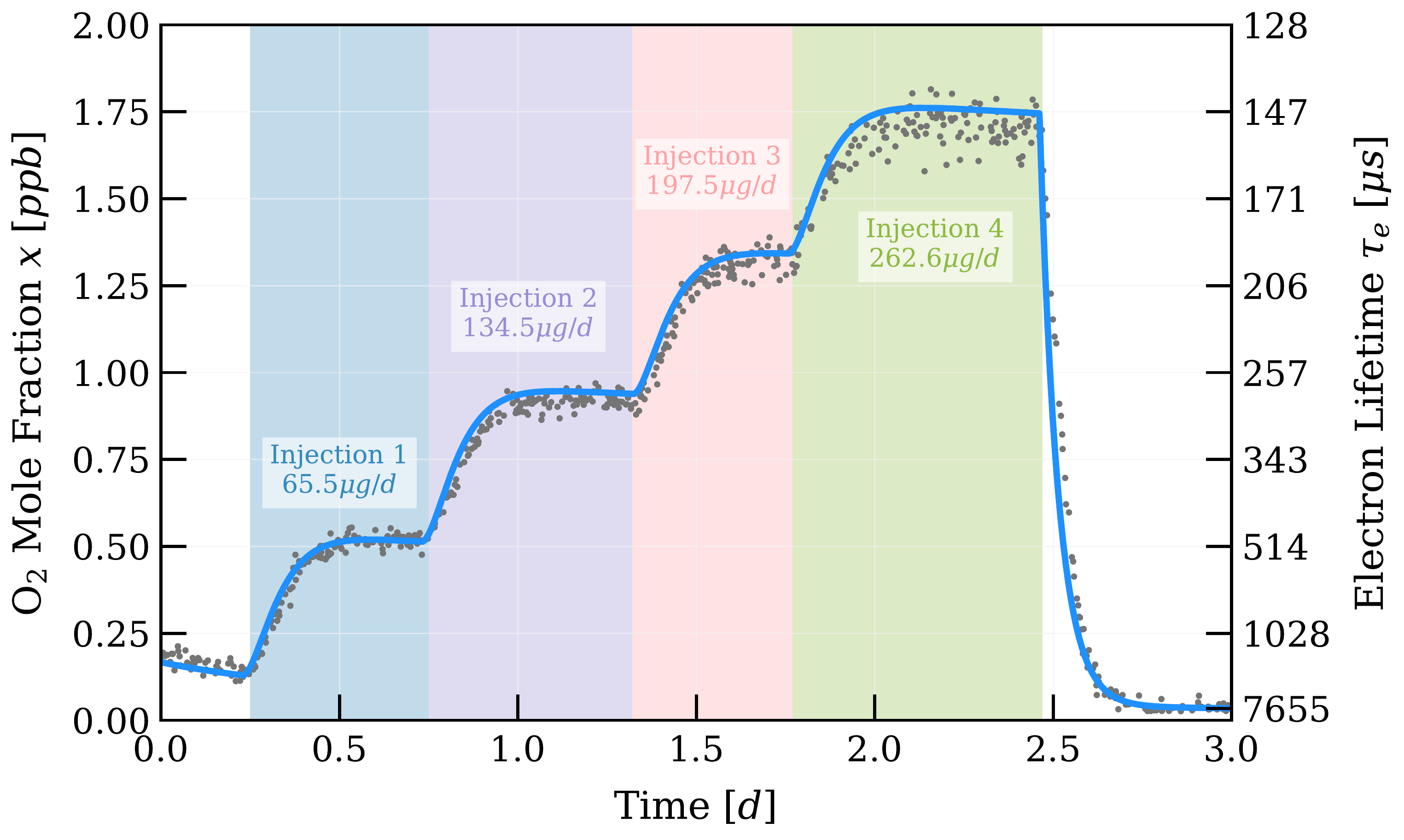}
\caption{Model (blue) of PM \otwo mole fraction \xpm (gray) during the injection sequence performed with copper-impregnated spheres in the \otwo filter vessel. The colored regions indicate the different injection rates. The time dependence of the \otwo source \Lpb is constrained by the equilibrium mole fractions before and after the injection sequence. The efficiency $\epsilon$ used is \epsilonQfive.}
\label{fig:q5fit}
\end{figure*} 

\subsection{Full Transport Model} \label{ssec:lxepur_model}
% LXe circuit flow without filter, modeling lifetime, effective ``distillation'' and new lambda. Point out this is similar to the getter bypass mode. This distillation efficiency of ~62\% was measured at multiple speeds.
Putting together these characterizations of circulation speed, impurity sources, and liquid-gas \otwo exchange, we can now construct a full model of \otwo transport in Xeclipse in order to extract the efficiency of the tested filters.
The flow configuration includes the circulation of liquid between the PM and CPS, with simultaneous continuous injection of \otwo via the GXe in the PM, and continuous purification of the pump buffer GXe volume through the getter.  Utilizing the elements of the generic model in Eq.~\ref{eq:generic_model}, this system can be described by

\begin{equation} \label{eq:cps_full_model}
\begin{split}
    \npm \frac{d \xpm}{dt} &= \npm \frac{f}{\Tcl} \big[(1 - \eff) \xpb - \xpm \big] + \Linj \text{ \hspace{40pt}  \big(PM\big)}, \\
    \npb \frac{d \xpb}{dt} &= \npm \frac{f}{\Tcl} (\xpm - \xpb) - \npb \frac{\xpb}{\Tdist} + \Lpb \text{\hspace{10pt} \big(CPS pump buffer\big)}.
\end{split}
\end{equation}

Here, \xpm and \xpb are the \otwo mole fractions in the PM LXe (measured) and the pump buffer LXe (not measured), respectively.
The amounts of LXe in the PM \npm, and in the pump buffer \npb are known from the two levelmeters.
The \otwo injection rate \Linj is known from the pressure in the \otwo injector reservoir volume (Sec.~\ref{ssec:o2inj}) and the recirculation time $\Tcl$ is measured.
The coefficient $f=\fval$ (Fig.~\ref{fig:fcalculation}) is used in both equations. 
The remaining unknown values are $\Tdist$, the sum of unknown \otwo sources entering the pump buffer LXe $\Lpb$ (generally time-dependent), and the efficiency of the filter $\eff$. 
The steady-state measured mole fraction $\xpm(t \gg \Tcl)$ can be derived (setting $\frac{d \xpm}{dt} = \frac{d \xpb}{dt} = 0$) as
\begin{equation} \label{eq:neq_full_model}
\xpm(t \gg \Tcl) = \frac{\Tcl}{f \npm} \bigg[ \Linj + \frac{\xi (1 - \eff)}{1 + \xi \eff} (\Lpb + \Linj) \bigg],
\end{equation}
where $\xi = \frac{f \Tdist}{\Tcl}$ determines the \otwo removal at the pump buffer liquid surface. For $\eff = 1$ (perfectly efficient filtration) or $\Tdist \ll \Tcl$ (extremely efficient removal via the GXe), the coefficient multiplying $(\Lpb + \Linj)$ in Eq.~\ref{eq:neq_full_model} goes to zero, and we are left with the equilibrium purity of Eq.~\ref{eq:asymptotic}.

\begin{figure*}[t]
\includegraphics[width=1.\textwidth]{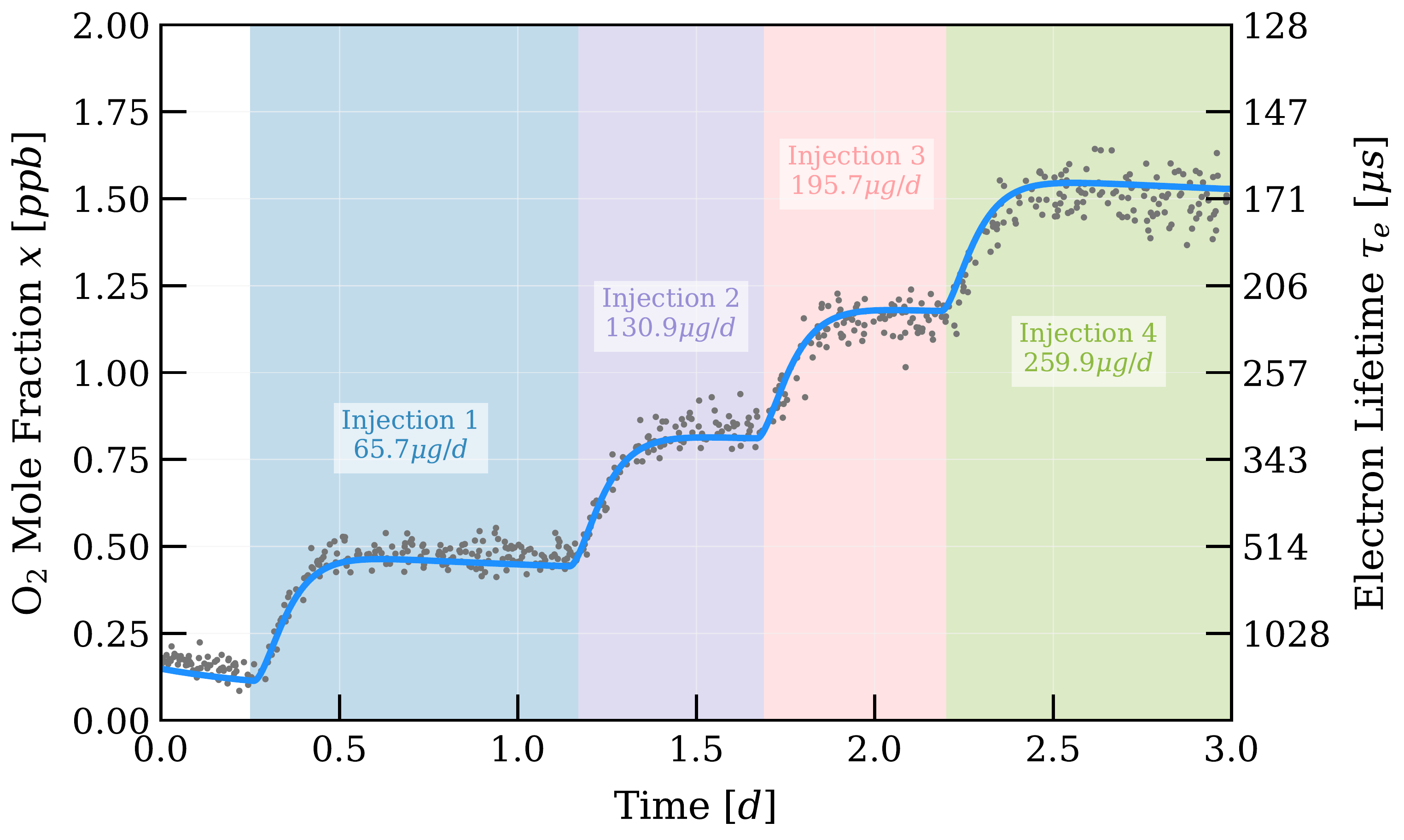}
\caption{Model (blue) of PM \otwo mole fraction \xpm (gray) during the injection sequence performed with non-evaporable getter pills in the \otwo filter vessel. The colored regions indicate the different injection rates. The time dependence of the \otwo source \Lpb is fixed by the equilibrium mole fractions before and after the injection sequence. The efficiency used is \epsilonAPI.}
\label{fig:apifit}
\end{figure*} 

%\subsubsection{Model Validation Procedure}
%\label{sssec:procedure}
A measurement procedure was developed to decouple the pump buffer \otwo source $\Lpb$ from the filter efficiency $\epsilon$ and the timescale of interphase transport in the pump buffer $\Tdist$. The LXe pump is started from a low purity near the measurement threshold of the PM, with purification of both GXe and evaporated LXe through the getter running in parallel. When an approximately constant purity is reached, an \otwo injection is begun by pressurizing the reservoir volume of the \otwo injector. The \otwo partial pressure is then increased several times, waiting each time until a new equilibrium is reached. Thus, several increasing values of $\xpm(t \gg \Tcl)$ in Eq.~\ref{eq:neq_full_model} are measured for known values of \Linj, producing a line whose slope is independent of $\Lpb$.

In reality, $\Lpb$ is time-dependent, complicating this relationship. We thus solve the differential equations in Eq.~\ref{eq:cps_full_model} numerically, 
assuming a decreasing $\Lpb$ as in Eq.~\ref{eq:lambdat}, with a small quadratic correction, such that the source reduces by half in about \sirange{0.5}{1.0}{\day}. 
The procedure still performs well at decoupling $\Lpb$ from the degenerate parameters, since this time-scale is much longer than $\Tcl$ and $\Tdist$ (both $< \SI{1}{\hour}$).

%\subsubsection{Measurement of $\tau_S$ for CPS Pump Buffer}
The timescale of interphase transport in the pump buffer was measured by replacing the filter vessel with an empty stainless steel tube and performing the above procedure. This is equivalent to setting $\eff = 0$ in Eq.~\ref{eq:neq_full_model}, giving

\begin{equation} 
\label{eq:nofilter}
\xpm(t \gg \Tcl) = \frac{\Tcl}{f n} \bigg[ \Linj + \xi (\Linj + \Lpb) \bigg].
\end{equation}

\noindent
The measured electron lifetime trend cannot be modeled without including the liquid-gas exchange of \otwo.
Eq.~\ref{eq:nofilter}~ is used to model the measured electron lifetimes, with a time-dependence for $\Lpb$ as in Eq.~\ref{eq:lambdat}. 
The result is shown in Fig.~\ref{fig:nofilterfit}, with $\hm = \PBhmeasured$ and $\Tdist = \PBtdistmeasured$.

This value is compared with the theoretical prediction using the procedure of Sec.~\ref{ssec:gas_exchange}, yielding values\\ $\hm = \PBhprediction$ and $\Tdist = \PBtdistcalcval$, demonstrating that interphase transport can explain the observed \otwo removal. The value $\Tdist = \PBtdistmeasured$ is used for all subsequent runs to evaluate $\epsilon$.

\section{Measurements of Filtration Efficiency} \label{sec:results}

With $\Tdist$ now fixed, the procedure above can be carried out to disentangle the remaining systematic $\Lpb$ from our parameter of interest $\epsilon$. 
The filter is installed, and circulation with the LXe pump proceeds until an equilibrium purity is reached. 
Then, injections at several fixed values of \Linj are performed, and the new equilibrium is measured. A time-dependent model is used to evaluate $\epsilon$ and $\Lpb(t)$.
During each of these tests, PM LXe evaporated through the heat exchanger is combined with GXe extracted from the CPS pump buffer and circulated through the getter in parallel. 
The effect of the traditional evaporated-LXe purification is fully fixed by the measured GXe flow and $f$. It also plays a very minor role in the dynamics compared to the far more efficient cryogenic purification, and thus is omitted from the equations of Sec.~\ref{ssec:lxepur_model} for simplicity.

\subsection{Copper-Impregnated Spheres (Engelhard Q5)}
% Q5 efficiency from rise and plateau of lifetime, and from multiple continuous injection rates.
The procedure described in Sec.~\ref{ssec:lxepur_model} was carried out for a filter vessel composed of a stainless steel tube with \SI{10.2}{\milli \meter} inner diameter and \SI{238}{\milli \meter} length, filled with \SI{18}{\gram} of Q5 contained between two \sfrac{1}{2} inch VCR snubber gaskets.
Beginning from a purity near the measurement threshold of the PM, the LXe pump is run until the electron lifetime begins to plateau. 
Then, \otwo is continuously injected at several fixed, known rates (shown in Fig.~\ref{fig:q5fit}) to evaluate the filter efficiency $\epsilon$ in Eq.~\ref{eq:cps_full_model}. Here $\xpm(t)$, \npm, \npb, \Tcl, and \Linj are measured; $f$ and \Tdist are fixed at \fval and \Tdistval respectively using the measurements described in Sec.~\ref{sec:model}, and \Lpb and $\epsilon$ are estimated by matching the data. 
The resulting model is shown in Fig.~\ref{fig:q5fit}. The rate of \otwo removal due to the purification of the pump buffer gas volume is fixed using $\hm = \SI{2.7e-4}{m/s}$ from the filter-less run. 
A $~t^{-1}$ time-dependence is used to model the \otwo introduction from the LXe pump, with a half-life decreasing from \SI{0.35}{\day} to \SI{0.83}{\day}. 
The value found for the Q5 filter efficiency $\epsilon$ is \epsilonQfive. Thus, at the liquid flow of \lxeflowQfive used in this test, the \otwo concentration in the xenon exiting the filter was \SI{8}{\percent} of the concentration at the inlet.

\subsection{Non-Evaporable Getter}
The same procedure was performed using a custom filter vessel from API~\cite{API}. 
The 
valved stainless steel vessel contains \apimass of NEG pills (\SI{4}{\milli \meter} diameter x \SI{3}{\milli \meter} thickness, \SI{0.196}{\gram} per pill) and is equipped with sintered discs at the inlet and outlet. 
This material is comparable to SAES St 707\texttrademark, the alloy used inside SAES high-temperature gas purifiers~\cite{st707}. 
The resulting model is shown in Fig.~\ref{fig:apifit}. The interphase transport rate in the pump buffer is again fixed at $\Tdist=\Tdistval$. 
A decreasing time-dependence (Eq.~\ref{eq:lambdat}) is used to model the \otwo introduction from the LXe pump, with half-life increasing from \SI{0.25}{\day} to \SI{0.55}{\day} over the five day range. 
The value used for the NEG filter efficiency $\epsilon$ is \epsilonAPI. Thus, at the liquid flow of \lxeflowAPI used in this test, the \otwo concentration in the xenon exiting the filter was \SI{34}{\percent} of the concentration at the inlet. 

\begin{table}[t]
\centering
\begin{tabular}{|c|c|}
\hline
Parameter Name & Value  \\
\hline\hline
    $f$ & \fval \\
    \hline
    $h$ & \PBhmeasured \\
    \hline
    $\epsilon$ (CIS) & \epsilonQfive \\
    \hline
    $\epsilon$ (NEG) & \epsilonAPI \\
    \hline
     
\end{tabular}
\caption{Table of key parameters evaluated in the Xeclipse system: the coefficient of purity equilibration $f$, the convection mass transfer coefficient $h$, and the efficiency of \otwo removal $\epsilon$ for the two filters, filled with copper-impregnated spheres (CIS) and non-evaporable getter (NEG) pills.}
\label{table:sherwoods}
\end{table}

\subsection{Discussion}
Both filter media demonstrated a significant removal rate of \otwo from the LXe stream at the flow tested. Injection rates as high as \SI{260}{\micro \gram \per \day} were successfully compensated by filtration, implying the viability of these filters for purification of multi-tonne LXeTPCs, with expected desorption rates $\mathcal{O}$(\sirange{10}{100}{\micro \gram \per \day}).
In both cases, the filter efficiency $\epsilon$ was found to be inconsistent with unity, meaning that the \otwo concentration at the filter outlet was not zero. 
Thus, a higher filter mass or reduced LXe flow would improve $\epsilon$, since both equivalently increase the contact time for the sorption reaction to take place. 
When varying the rotation frequency of the LXe pump, the pump has a much higher tendency to become gas-bound, so no systematic modulation of flow speed was permitted, but tests confirmed that the filtration efficiency decreases with increasing flow between \sirange{0.1}{0.2}{\liter \per \minute}. 
Therefore, at the $\approx \SI{2}{\liter \per \minute}$ LXe flow speed required for multi-tonne experiments, maintaining similar filtration efficiency likely requires an increase in the filter mass by a similar factor. 
For the copper-impregnated spheres, this would likely amount to a \rn emanation rate too large for the experimental goals of XENONnT~\cite{xenonnt}, though the efficiency would again be close to unity.
The NEG, however, falls easily within the radiopurity requirements of the experiment, even with such an increase in filter mass.
An approach combining an initial copper-impregnated sphere filtration followed by a change to NEG filtration is also viable, since the high purification speed allowed with the former can bring many tons of LXe to a high purity in a short time, and the resulting \rn would decay away after the switch.

The copper-impregnated spheres were successfully regenerated before use, and the NEG filter was successfully reactivated by heating and pumping. 
Both filters tested successfully captured $\mathcal{O}$(\sirange{1}{10}{\milli \gram}) without showing any decrease in efficiency, further confirming their viability for long-term operation of multi-tonne LXeTPCs without requiring frequent reactivation. 

\section{Conclusion}
We have designed and tested filters that can efficiently purify LXe to the level of $\mathcal{O}(\SI{10}{ppt})$, and which can be adapted for application to multi-tonne LXeTPCs while maintaining the extremely stringent radiopurity required for dark matter direct detection. 
The two filter media tested, copper-impreg\-nated spheres and NEG pills, both captured electronegative impurities at the rate required to offset sources of \otwo in such detectors, and could be successfully regenerated for long-term use. 

\begin{acknowledgement}
We gratefully acknowledge the continued support from the National Science Foundation for the XENON project at Columbia University, which enabled this work. We also thank Christian Weinheimer for lending the cryogenic liquid pump used in Xeclipse and Masaki Yamashita for suggesting and providing the NEG tested in this work. Finally, we thank Masaki Yamashita and Masatoshi Kobayashi for many useful discussions and Michael Murra for his careful reading of the manuscript.
\end{acknowledgement}

\bibliographystyle{spphys}       % APS-like style for physics
\bibliography{bibliography}  % Produces the bibliography via BibTeX.

\end{document}